\documentclass[aps,prl,twocolumn]{revtex4-1}
\usepackage{graphicx, epstopdf}

\begin{document}


\title{Growth and Oxygen Doping of Thin Film FeTe by Molecular Beam Epitaxy}


\author{Mao Zheng, Hefei Hu, Can Zhang, Brian Mulcahy, Jianmin Zuo and James Eckstein}
\email[]{eckstein@illinois.edu}
\affiliation{University of Illinois Urbana-Chamapaign}


\date{\today}

\begin{abstract}
FeTe is known to become a superconductor when doped with oxygen.  Using layer-by-layer growth of single crystal films by molecular beam epitaxy (MBE), we have studied how oxygen incorporates.  If oxygen is supplied during growth of a layer, it substitutes for tellurium inhomogeneously in oxygen domains that are not associated with superconductivity.  When oxygen is supplied after growth, it diffuses into the crystalline film and incorporates interstitially.  Only the interstitial oxygen causes superconductivity to emerge. This suggests that the superconductivity observed in this material is spatially uniform and not filamentary.
\end{abstract}


\maketitle

\section{Introduction}
Due to  their unusual composition, multi-band nature\cite{pp-orbital} as well as potential technical applications\cite{qianglis-fesc-review}, iron base superconductors have captured the attention of the research community since their discovery\cite{discovery-fesc}. In particular, the “11” system, of Fe chalcogenide with PbO structure, has garnered interest because of its simple structure, relatively mild toxicity and rich array of properties accessible through doping\cite{fesc-doping1},\cite{s-dope} and stoichiometric\cite{fesc-fe-sensitivity} variations. Optimally doped FeTe$_{1-x}$Se$_{x}$ (x=0.5) samples have $T_c$=$14\,^{\circ}{\rm K}$ at atmospheric pressure\cite{qianglis-fesc-review} and T$_c$ of FeSe samples under pressure can go as high as $30\,^{\circ}{\rm K}$ \cite{p-induced-sc}. The parent compound of the ‘11’ system, FeTe, itself is not a superconductor. FeTe undergoes a insulator-metal transition around $70\,^{\circ}{\rm K}$, coinciding with a lattice structural transition from high temperature tetragonal to low temperature monoclinic symmetry and the establishment of collinear antiferromagnetic order\cite{phase-tran-near-70K}. 

\par In thin film form, FeTe can incorporate oxygen and become superconducting\cite{shi-fete-oxygen-pld},\cite{nie-fete-anneal}. There are two possible ways for oxygen to go into a FeTe film. Oxygen atoms can covalently bond to Fe thus substituting for Te in the lattice, or they can be incorporated as interstitials. In previous studies, Si, et. al., observed a superconducting transition in FeTe films grown in an oxygen environment and postulated substitution as a possible incorporation mechanism\cite{shi-fete-oxygen-pld}. On the other hand, Nie et al found that superconductivity could be reversibly introduced via annealing in an oxygen environment\cite{nie-fete-anneal}. This evidence points to oxygen interstitials being responsible for superconductivity. Since interstitial oxygen is not covalently bonded to the lattice, it should be much more mobile than substitutional oxygen and the kinetics of its incorporation should reflect this.  Here we report that while both substitutional and interstitial oxygen can occur in oxygen-doped FeTe films, the balance between the two depends on how the oxygen doping is carried out. The substitutional oxygen incorporates inhomogeneously in domains while the interstitial oxygen is homogeneously distributed. Furthermore, substitutional oxygen does not cause superconductivity, rather it is the interstitial oxygen that is responsible for this change. These findings suggests that superconductivity in oxygen doped FeTe is spatially homogeneous and not due to filamentary coupling of oxygen substituted domains.

\section{Film Growth and Characterization}
FeTe films were grown in a custom designed MBE system with base pressure of $1 \times 10^{-9}$ Torr\cite{ecksteinmbebookchapter}. Elemental charges of Fe and Te were evaporated from individually shuttered effusion cells onto heated LaAlO$_3$ (LAO) substrates. The growth process was monitored using Reflection High Energy Electron Diffraction (RHEED). A substrate temperature of $300\,^{\circ}{\rm C}$ was found to be optimum after a series of test growths between $250\,^{\circ}{\rm C}$ and $440\,^{\circ}{\rm C}$. Temperatures much lower than $300\,^{\circ}{\rm C}$ are insufficient for FeTe crystallization while higher temperatures promote the $\langle101\rangle$ orientation, similar to what has been reported for FeSe MBE growth\cite{fese-mbe}. Around $300\,^{\circ}{\rm C}$, the surface evaporation rate of extra Te is large enough to lead to self-regulated Te incorporation analogous to MBE growth of III-V semiconductors. Because of this, the growth rate was determined by the Fe flux. For films in this study, a Te:Fe flux ratio of approximately 2:1 was used with a deposition rate of 0.1\AA/s. While the Te shutter stayed open during the growth, the Fe shutter was closed for a 15 second anneal period after each layer. Rutherford backscattering (RBS) analysis on a calibration film grown under these conditions showed that the composition ratio was Te:Fe=1:1.09, with $\pm0.05$ uncertainty, similar to that found in bulk samples\cite{fete-bulk-lattice}.

%

 \begin{figure}
 \includegraphics{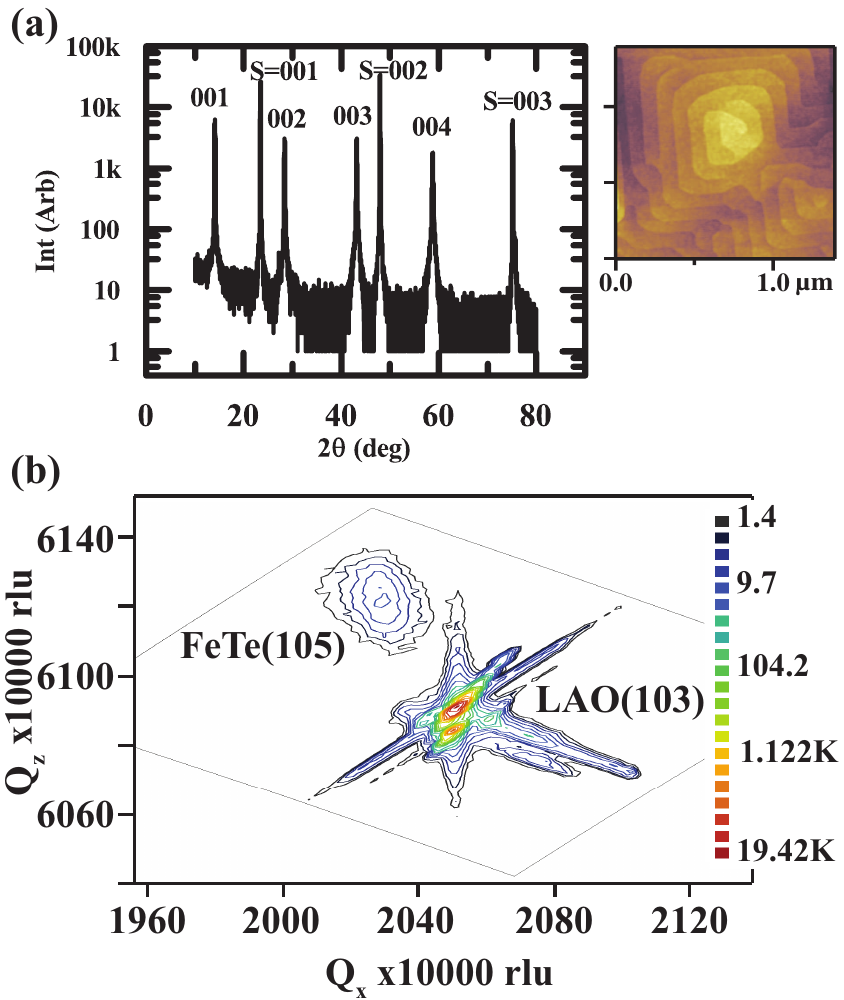}
 \label{xrdrsm}
 \caption{(a) XRD pattern of FeTe film on LAO substrate, indicating single phase and single orientation growth. Substrate peaks are labeled "S". Inset is an AFM image showing atomic terraces and screw dislocations, maximum height variation is 3.5nm over an area larger than $1\mu$$m^2$ (b) RSM around LAO(103) and FeTe(105) peak. The film is relaxed to bulk value, with $a=0.383$nm and $c=0.6275$nm}
 \end{figure}

Figure 1(a) shows an x-ray 2$\theta-\omega$ diffraction (XRD) pattern of a single crystal FeTe film grown on a LAO substrate. The film is (001) oriented, and no second phase or mis-orientation defect can be seen. Figure 1(b) is a reciprocal space map (RSM) near the FeTe(105) and LAO(103) diffraction peaks.  The in-plane lattice spacing of the substrate and the film are different, showing that the film is not clamped to the substrate lattice.  In fact the FeTe films have in-plane lattice spacing similar to unstained bulk samples\cite{fete-bulk-lattice}.  The inset to Figure 1(a) shows an atomic force microscopy (AFM) image of a typical FeTe film surface. Both screw dislocations and atomic terraces on the order of 100nm can be seen.

 \begin{figure}
 \includegraphics{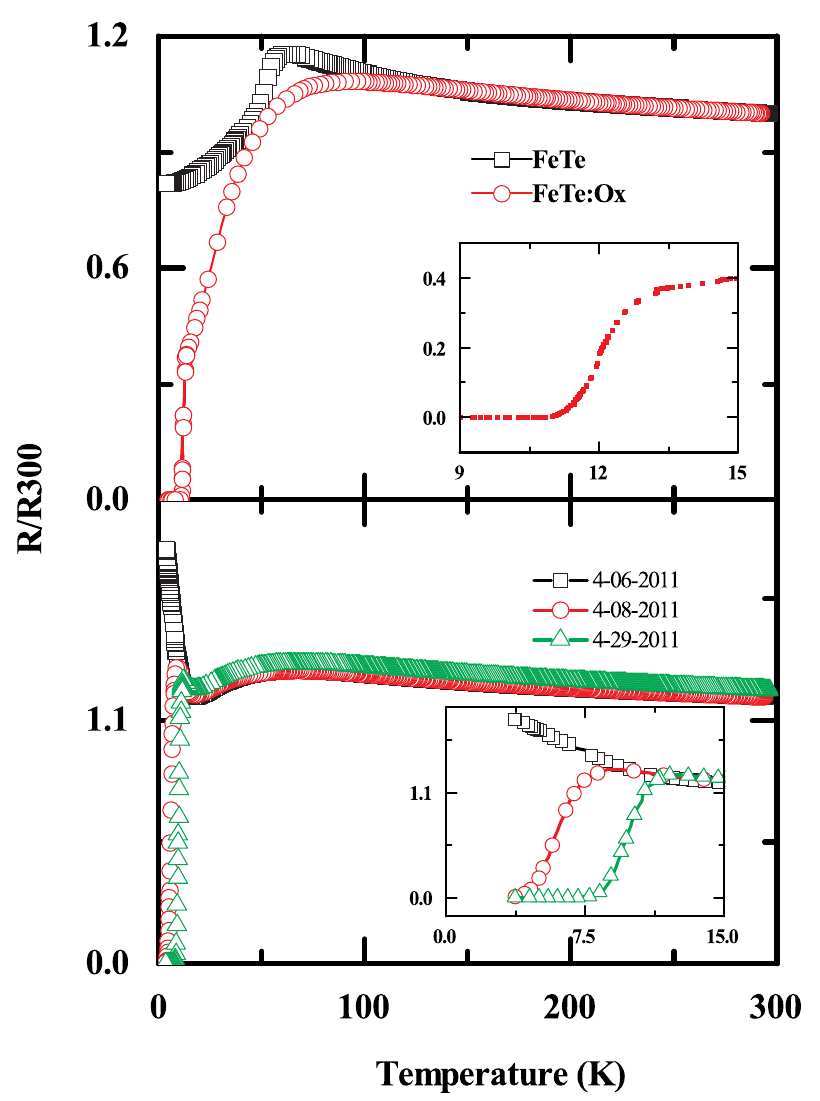}
 \label{RTs}
 \caption{(a) Open square curve shows normalized resistance for typical FeTe grown without oxygen. Open circle curve shows as-grown oxygen doped FeTe film with superconducting transition. Inset shows a zoom-in of R(T) of oxygen doped FeTe near T$_c$. (b) Emergence of superconductivity in an FeTe film grown without oxygen before and after expose to air, inset shows a zoom-in near the transition region. The curve with the lower T$_c$ value was measured 2 days after growth and the curve with higher T$_c$ value was measured 21 days later}
\end{figure}

Undoped FeTe films fresh out of the chamber were not superconducting. A resistance versus temperature, R(T), in-plane transport measurement for a typical undoped FeTe film is shown by the open square curve in Figure 2(a). The insulator to metal transition near $70\,^{\circ}{\rm K}$, which is known to associate with an antiferromagnetic and structural phase transition\cite{phase-tran-near-70K}, can be clearly observed. Once such films were exposed to air, superconductivity gradually developed. Figure 2(b) shows the emergence of superconductivity over a 23 day period in an undoped FeTe film stored in air.

\par Superconductivity can also be obtained in FeTe films by doping with oxygen \emph{in-situ}. To study how this occurs, we introduced oxygen into FeTe in two different ways. In the first, films were grown epitaxially at $300\,^{\circ}{\rm C}$ and cooled down to below $100\,^{\circ}{\rm C}$ always in an oxygen beam (4 to 8 $\times 10^{15} cm^{-2}sec^{-1}$ oxygen flux). In the second, FeTe films were grown epitaxially in vacuum, and the oxygen beam was only turned on during the cooling stage when the film temperature was between $300\,^{\circ}{\rm C}$ and $100\,^{\circ}{\rm C}$. Both methods produced single crystal films with sharp superconducting transitions. Typically, $T_{c}^{onset} = 12.5\,^{\circ}{\rm K}$ with transition width around $1.5\,^{\circ}{\rm K}$.  The open circle curve in Figure 2(a) shows R(T) for one of the oxygen doped films. As we will show, oxygen supplied during growth is incorporated substitutionally for covalent Te, while oxygen supplied during cooling is incorporated as interstitially.

\section{Microscopy and Spectroscopy}

 \begin{figure}
 \includegraphics{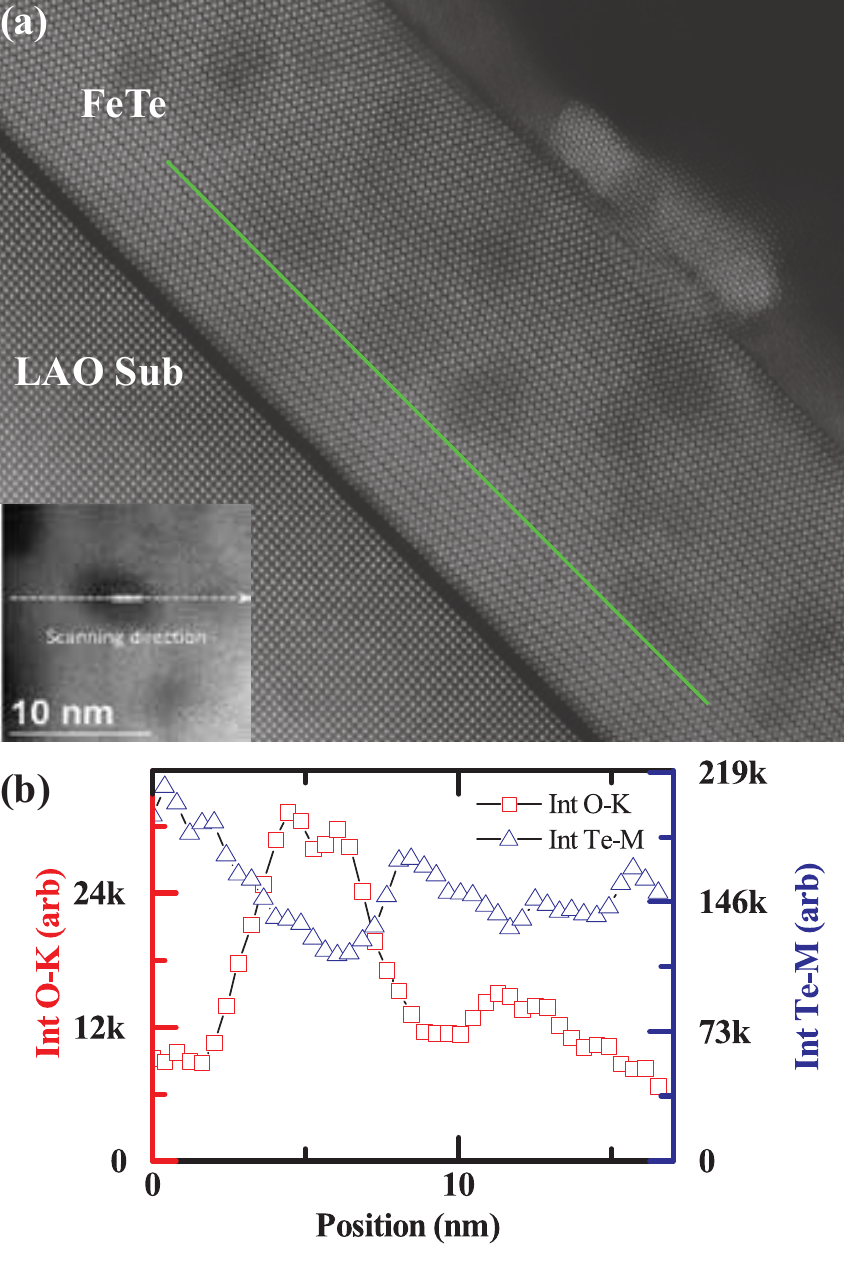}
 \label{stem-eels}
 \caption{(a) STEM cross-sectional view of FeTe film grown and cooled down in oxygen. Green line denotes the 8$^{th}$ layer, where oxygen beam was turned on. The dark regions after the 8$^{th}$ layer are oxygen substitution regions. Inset shows a line scan trajectory over a dark region. The length bar is the same in both images (b) Integrated EELS signals for O (square) and Te (triangle). An oxygen concentration peak coincides with a Te concentration dip around 5nm.}
 \end{figure}

To study the mechanism of oxygen incorporation, an FeTe film grown partially in an oxygen beam was imaged using Scanning Transmission Electron Microscopy (STEM). Figure 3(a) shows a High Angle Annular Dark-Field (HAADF) cross-sectional image of the sample. The green line in the figure marks the 8$^{th}$ layer of FeTe at which point the oxygen beam was turned on, i.e. layers prior to that were grown in vacuum and layers after that were grown in oxygen. After growth, the film was cooled in oxygen and was superconducting at $12.5\,^{\circ}{\rm K}$ out of the chamber. In the layers grown in oxygen (above the green line in the figure), spherical dark regions about 4 nm in diameter can be seen. Since the contrast (brightness) of an atomic column in dark-field STEM is proportional to the square of the atomic number ($Z^2$) of atoms in the column, the dark regions contain lighter elements compared to the rest of the film. While the FeTe single crystal structure still surrounds the dark region, an Electron Energy Loss Spectroscopy (EELS) line scan across one of these dark regions reveals an increase in oxygen content at the expense of Te.  Figure 3(b) shows energy integrated O-K edge and Te-M$_{4,5}$ edge signals along the line scan. First, inside the dark region, a prominent peak of oxygen, which coincides with a dip in Te, can be seen. This suggests that some Te is substituted by oxygen in the dark region. Second, outside the dark region, a relatively flat but non-zero oxygen signal is detected along with a roughly constant Te signal. This indicates a homogenous oxygen concentration in the rest of the film.
\par If we assume that the EELS signal is proportional to atomic concentration by ignoring higher order scatterings and also assume that each dark region contains one oxygen-rich sphere, we can estimate the concentration of oxygen substitution in the dark regions and the interstitial concentration in the rest of the film. Although the exact thickness of STEM samples is not precisely known, we estimate it to be 20nm thick, as typical STEM samples range between 10nm and 30nm.  Since the Te curve is reduced by about 20\% in Figure 3(b) and the oxygen substituted defect sphere is about 20\% of the sample thickness, virtually all of the Te must be substituted by oxygen within the sphere. Similarly, the homogenous baseline of the O-K edge, which is due to uniformly distributed oxygen interstitials, is about 20\% of the peak.  If we scale for the 20\% relative size of the defect sphere in the total thickness of the sample, we get the oxygen interstitial concentration to be about 4\%.  Since neither STEM nor XRD detected the presence of a second phase, the FeO$_x$Te$_{1-x}$ from oxygen substitution must have incorporated epitaxially with the same symmetry as the  surrounding FeTe. 
\par In the initial layers grown before the oxygen beam was turned on, the dark oxygen substituted regions are absent. Any oxygen in these layers was incorporated through diffusion after the layers were grown and must be interstitial.  Furthermore, similar films grown entirely in vacuum but cooled in oxygen are superconducting with sharp transitions.  Thus, interstitial oxygen alone is sufficient to cause superconductivity.
\par While this STEM-EELS analysis shows that a sample grown and cooled in oxygen contains both substitutional as well as interstitial oxygen, and we know that substitutional oxygen is formed only during the growth, it is still unclear whether substituted oxygen alone can cause superconductivity. To distinguish contributions, if any, to superconductivity from substitutional and interstitial oxygen, and to probe the condition under which interstitial oxygen is incorporated, another film was grown in oxygen and immediately capped with a thick aluminum oxide layer before cooling down in oxygen. The purpose of the aluminum oxide cap was to provide a barrier to preserve the oxygen concentration formed during growth.  As shown in Figure 4 open square curve, this film did not superconduct, and interestingly it remained unchanged even after prolonged exposure to air. Unlike uncapped films which eventually became superconducting.

 \begin{figure}
 \includegraphics{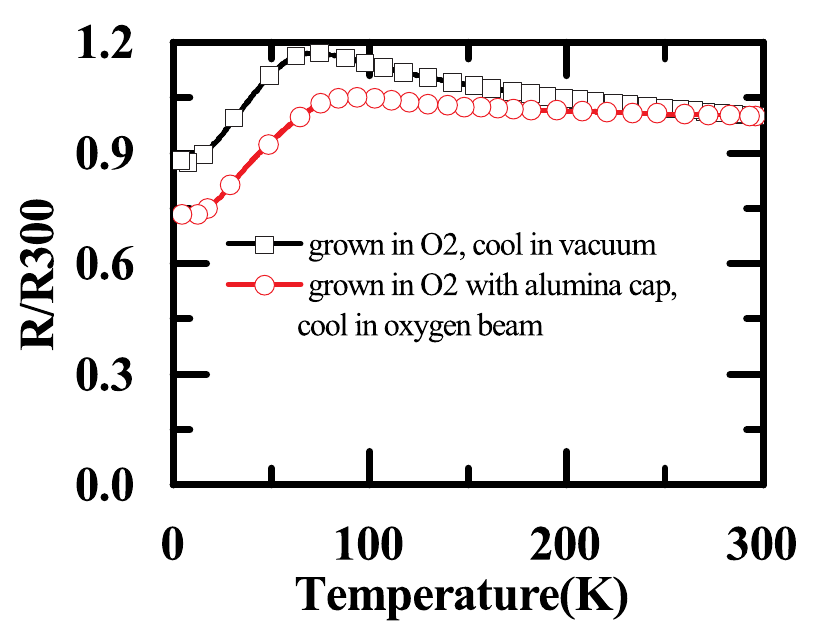}
 \label{diff_doping}
 \caption{Normalized resistance for two films. The open square curve shows a film grew in oxygen beam but cooled down in vacuum. The film was not superconducting fresh out of the chamber, but turned superconducting after exposure to air. The open circle film was grown in oxygen beam and capped with a gas impervious alumina cap and cooled down in oxygen. This film remains non-superconducting.}
 \end{figure}

\par From this experiment, we can draw three important conclusions. First, the persistent absence of superconductivity confirms the effectiveness of the aluminum oxide cap as a diffusion barrier.  Second, the absence of superconductivity, combined with the presence of a effective diffusion barrier, indicates that the degree of interstitial oxygen incorporation during growth at $300\,^{\circ}{\rm C}$ is insufficient to lead to superconductivity. Third, based on our STEM-EELS study, this film must contain substitutional oxygen since it was grown in the presence of oxygen. From this we can conclude that substitutional oxygen alone does not lead to superconductivity. Rather, the oxygen that gives rise to superconductivity is interstitial and is incorporated from post-growth oxygen exposure. To verify that substitutional oxygen does not lead to superconductivity, but interstitial oxygen does, another film was grown under identical oxygen pressure but cooled down in vacuum. Although non-superconducting initially, as shown by Figure 4 open square curve, it became superconducting after exposure to air.

\section{Conclusions}
In conclusion, we have studied how oxygen incorporates into FeTe thin films, causing them to become superconductors. The films have $T_{c}^{onset}\sim12.5\,^{\circ}{\rm K}$ and transition width around $1.5\,^{\circ}{\rm K}$. By systematically using STEM-EELS and a comparison of transport in films grown under different conditions, we conclude that oxygen incorporated during growth substitutes for Te inhomogeneously with isolated regions of the film almost entirely made up of oxygen substituted FeO$_x$Te$_{1-x}$. It is remarkable that this occurs without disrupting the crystalline long range order. On the other hand, oxygen incorporated after the film growth forms a  homogeneous interstitial distribution. The substitutional oxygen alone does not lead to superconductivity, while the interstitial oxygen does. This suggests that the superconductivity observed in oxygen doped FeTe films is homogenous and not due to a filamentary connection between oxygen substituted domains. Furthermore, insufficient interstitial oxygen is present at the growth temperature of $300\,^{\circ}{\rm C}$ to lead to superconductivity. Interstitial oxygen is predominately incorporated during exposure at lower temperatures via diffusion through the film.

\begin{acknowledgments}
This work is supported as part of the Center for Emergent Superconductivity, an Energy Frontier Research Center funded by the US Department of Energy, Office of Science, Office of Basic Energy Sciences under Award No. DE-AC0298CH1088. Film characterization was carried out in part in the Frederick Seitz Materials Research Laboratory Central Facilities, University of Illinois.
\end{acknowledgments}

\bibliography{FeTe_paper_play}

\end{document}